\documentclass[12pt,preprint]{aastex}

\slugcomment{ }

\shorttitle{Dynamically stable CVs}
\shortauthors{Bianchini et al. 2006}

\begin{document}



\title{ Do magnetic fields contribute to 
the dynamical stability of the secondaries of CVs  against
mass transfer?
 }

\author{A. Bianchini$^{1,2}$, F. Tamburini$^{2}$, P. Johnson$^{1}$}

\altaffiltext{1}{Department of Physics and Astronomy, University of Wyoming,
Laramie, WY 82071.} \altaffiltext{2}{Department of Astronomy, 
University of Padova, Italy.}
\begin{abstract}
We show  the presence of  CVs  with  orbital periods in the range 3.5-7 h 
and  mass ratios $q=M_2/M_1$ above the critical values for 
dynamically stable mass transfer.
We explore whether  the magnetic fields produced in the 
secondaries  alter the theoretical mass ratio 
limits allowing dynamical stability.
Since magnetic fields  change the 
specific heats and the polytropic exponent $\gamma$ 
of the gas in the convective envelope, the mass-radius  
adiabatic exponent $\xi_{ad}$  also changes.
We find that  turbulent magnetic fields can produce 10\%
less restrictive critical q-profiles while  large-scale toroidal
and poloidal fields have smaller effects. Thus, magnetic fields alone
do not  account for the stability of  all the anomalous CVs.  
However, we found that the small variations of $\xi_{ad}$ induced by  
solar-type magnetic cycles  explain the amplitudes of the cyclic accretion 
luminosity variations shown by several CVs.  
\end{abstract}


\section{Introduction}

Cataclysmic Variables (CVs)  are semidetached  binary systems 
formed by a white dwarf (WD) primary accreting 
matter from a Roche lobe-filling secondary, usually a low mass main 
sequence star. 
The WD  is surrounded by an accretion disk, unless 
its magnetic field is strong enough to partially or totally 
control the accretion geometry; these two cases correspond to the 
{\it Intermediate Polar} systems  and the 
{\it Polar} systems, respectively.

CVs  are thought to lose angular momentum through
magnetic braking and gravitational radiation.
The shrinking of the Roche lobe of the secondary produces the 
mass overflow through the inner Lagrangian point and accretion 
onto the WD. Thus, CVs are generally evolving towards shorter orbital 
periods and less massive secondaries. 
The luminosity of a CV is mainly produced by the thermalization of 
the potential energy  released by the accretion stream.

Very few objects have been observed at orbital periods in the range
2-3 h, called the period gap (PG). This is   generally 
explained by the disrupted magnetic braking model (Rappaport, Verbunt 
\& Joss 1983; Spruit \& Ritter 1983).  When 
the orbital period gets close to 3 h, the mass losing secondary 
becomes fully convective and the  efficiency of the magnetic braking
suddenly decreases. 
As the mass transfer rate drops, the  secondary tends to recover 
its thermal equilibrium shrinking inside its Roche lobe and 
the CV is  {\it switched off}. 
Mass transfer would then  start  again only when angular
momentum losses due to gravitational radiation have brought the secondary
back in contact with its Roche lobe, which  typically occurs 
at approximately 2 h of orbital period. 

\smallskip
Not only does the history of the mass transfer rate of a CV 
characterize its secular evolution,
but it also characterizes the behavior of some subclasses of CVs. 
For example, dwarf novae (DN)  have  relatively low accretion 
rates so their  disks can be  thermally unstable and produce  
cyclic instability  phenomena.
Old novae and nova-like systems, instead,  are systematically 
brighter than DN  suggesting that their accretion disks are
powered by larger mass transfer rates. In this case, the higher 
temperatures of the accretion disks would 
in most cases prevent the onset of the instability phenomenon.
However, the most important difference between novae and dwarf novae 
is represented by the requirement that the primaries of  nova systems 
must have masses  larger than $\sim 0.5 M_{\odot}$ in order to produce 
ejecta (Livio 1992). 
The spread of the observed mass transfer rates in CVs  and, perhaps,
also the existence of various subclasses at any given orbital period, 
might have several causes.  The presence of magnetic cycles 
of activity in the secondaries (Bianchini 1987, 1990; Warner 1988; 
Richman, Applegate \& Patterson 1994) account only for 
luminosity changes of a few tenths of a magnitude.
More important mechanisms could be a variable efficiency of the envelope 
chaotic  dynamo combined with that  of the boundary layer dynamo 
(Zangrilli, Tout \& Bianchini 1997), secular mass transfer cycles 
excited by the irradiation of the secondary by the primary 
(King et al. 1996), and/or the hibernation scenario
for classical novae proposed by Shara et al. (1986). 
\smallskip

A CV emerging from common envelope (CE) evolution (Paczynski 1976) must
experience stable mass transfer from the Roche lobe filling secondary
on to the WD primary.
As continuous mass loss  perturbs the thermal equilibrium of the 
secondaries, they  are slightly smaller or larger than MS 
stars when they are predominantly  radiative or convective, 
respectively (e.g. Whyte \& Eggleton 1980, Stehle, Ritter \& Kolbe 1996, 
Baraffe \& Kolbe 2000).
Generally speaking, the stability of mass transfer is  
determined by the  change in radius of the secondary due to 
adjustment of its convective outer layers
(dynamical stability) and internal structure (thermal stability) in 
response to mass loss,  compared with the
changes of the Roche lobe radius induced by both the mass exchange and 
the  angular momentum losses by the system.
We define three exponents, $\xi_{ad}$, $\xi_{the}$ 
and $\xi_{RL2}$, for the adiabatic, thermal equilibrium and Roche-lobe 
mass-radius relations of low-mass MS stars, respectively. 
Since the time derivative of the radius can be 
written in the form
\begin{equation}
\frac{\dot R}{R}= \xi\frac{\dot M_2}{M_2}
\end{equation}
the binary will be adiabatically and thermally stable against mass 
transfer only if we have 
\begin{equation}
\xi_{ad}~-~\xi_{RL2} > 0
\end{equation}
 and
\begin{equation}
\xi_{the}~-~\xi_{RL2} > 0
\end{equation}
respectively.
Alternatively,   a second CE episode would quickly drive the two 
components of the newborn CV to cohalescence.
The $M_2-\xi_{ad}$ and $M_2-\xi_{the}$  profiles for stable
mass transfer were discussed by Schenker, Kolb \& Ritter (1998).
They can be easily transformed into two relations between  
$M2$ and the critical mass ratio $q_{critic}=(M_2/M_1)_{critic}$ 
as described by de Kool (1992) and King et al. (1994).
The critical q-profiles for adiabatic and thermal equilibrium
stable mass transfer can be described as follows.
For secondaries below $\sim$0.45 $M_\odot$, the deep convective 
envelope  tends to expand adiabatically in response to  mass loss 
on a dynamical time scale.
In this case, the dynamical stability criterion is the most 
restrictive one and the mass ratios  should satisfy the condition
$q<q_{dyn}=0.63$. 
For masses above $\sim$0.75$M_\odot$, the mass in the convective 
envelope is too small to produce adiabatic expansion, and the secondary 
even tends to shrink. As a consequence, dynamical stability is
guaranteed even for larger mass ratios while the     
thermal stability criterion $q<q_{therm}=1.25$ suddenly 
becomes the more restrictive one. 
All CVs are then expected to have mass ratios  
that are roughly  consistent with  these  stability criteria 
although  exceptions are possible (see next section 2). 
\bigskip

In this paper we explore  the effects of  magnetic fields 
on the critical  $q_{dyn}$ profile of lower mass
secondaries.
In section 2 we present   systems that apparently 
deviate from the  theoretical  stability criteria.
In section 3 we  try to evaluate how  magnetic fields  modify 
the critical $q_{dyn}$  profile.
The effects of the presence of solar-type  magnetic cycles on the mass 
transfer rate  is discussed in  section 4.    
Conclusions  are given in section 5.


\section{The q-profile of the observed CVs}
The   Ritter \& Kolb (2005) on-line  catalogue 
(hereafter RK2005) reports the masses and the mass ratios for 
about 102 CVs. 
The mass transfer stability criteria described by  de Kool (1992), 
 King et al. (1994)  and Shenker, Kolb \& Ritter (1998)
roughly state that, for mass companions below 0.45$M_\odot$, 
the dynamical stability criterion is the most
restrictive one and it must be $q<0.63$, while, for masses above 
0.75$M_\odot$, as the  dynamical stability is achieved even for 
very large mass ratios, the prescription for  thermal 
stability $q<1.25$  dominates.  
The  critical  $q(M_2)$ values for the dynamically and 
thermally  unstable mass transfer are sketched in the $q-M_2$ 
plane of Fig. 1 by the solid line ($q_{dyn}$) and the dashed 
line ($q_{therm}$), respectively.
The $M_2$ and  $q$ values of the selected sample of CVs 
are plotted in Fig. 1 using different symbols for different subclasses.
Polar and Intermediate Polar systems  are circled 
by open circles and open squares, respectively.
Looking at Fig. 1, we first notice that  $M_2$ and $q$  are 
linearly correlated, as predicted by standard CV models.  
We also observe that some of the
data points are not below the critical $q_{dyn}$ 
line  as requested by the stability criterion.  
In fact, six  CVs fall  inside  the thermally stable but 
dynamically unstable region; they are V1043 Cen, LX Ser, HY Eri, AT Ara, 
CM Del and V347 Pup. We should have also included  UX UMa 
for which RK2005 report  $M_1=M_2=0.47 M_\odot$ (Baptista et al. 1995).
However, from new radial velocity curves Putte et al. (2003)  derived 
$M_1=0.78\pm 0.13 M_\odot$ and $q=0.60\pm0.09$ that places UX UMa
just below the limit  for mass transfer stability.
Five   objects almost  coincide with the $q_{dyn}$ line; they are 
the old novae T Aur, HR Del, and DQ Her, the magnetic system V1309 Ori
and the nova-like RW Tri. 
Finally,  the recurrent nova CI Aql (Lederle \& Kimeswenger 2003)  
and the dwarf nova EY Cyg (Tovmassian et al. 2002) even  appear both 
thermally and dynamically unstable, but having rather long orbital 
periods they probably host evolved secondaries. In particular,
G\"ansicke et al. (2003) suggested that EY Cyg might have passed 
through a phase of (unstable) thermal timescale mass transfer
that produced a CNO-enriched secondary stripped of its outer layers.
Pollution of the secondary of CY Cyg  caused by an unrecorded nova 
explosion was also suggested by Sion et al. (2004).
Somewhat evolved secondaries might  also exist in short orbital 
period systems and, in any case, secondaries are mass losing stars
out of thermal equilibrium. Thus, the theoretical prescriptions 
for the dynamical stability of actual CVs  are probably different 
from those predicted for MS secondaries.
In this paper we  suggest that  the critical q-profile could be also 
modified by the magnetic fields produced in the convective envelopes of 
the secondaries.
\smallskip

Fig. 2  plots   $M_2$ {\it versus} $M_1$.
Filled circles represent masses that have been determined
spectroscopically; open circles, instead, refer to mass estimates
obtained from empirical correlations. These two classes of 
data points are indistinguishable in the diagram.
The errorbars of the suspected unstable objects are shown.   
Only V1043 Cen and LX Ser can be considered unstable at a 95$\%$ 
confidence level.
The six systems that fall inside the dynamically unstable region are
characterized by $M_{2(mean)}=0.38 \pm 0.1 M_\odot$ and 
$M_{1(mean)}=0.55\pm 0.1 M_\odot$. 
We  notice that these 'unstable objects' are dynamically  but not
thermally unstable (see Fig. 2).  The only exception is the peculiar 
dwarf nova EY Cyg discussed above. 
The  weak correlation between the  masses of the 
two components observed in Fig 2 is probably only due to the 
selection effects introduced by the thermal stability criterion.
Fig 2  suggests that  apparently unstable CVs tend to cluster around 
massess of the secondaries that correspond to orbital periods above 
the period gap.
Since the masses of the primaries are  very weakly correlated with 
the orbital period because they have  only to obey to the mass ratio
limits for stability, unstable systems must also  select the less 
massive primaries.
\smallskip

In order to investigate the  nature of these apparently dynamically 
unstable CVs, we plot in Fig. 3 the mass-period diagram.
Different symbols are like in Fig. 2 and basically show the same 
statistics.  
Empirical mass-period relationships were given, amongst others, 
by Warner (1995a,b),  Smith \& Dhillon (1998), Howell \& Skidmore (2000),  
and Patterson et al. (2003). 
Smith \& Dhillon (1998) found that the secondary stars in CVs with 
periods below 7-8 h are  indistinguishable from main-sequence stars.
As an example, we show in Fig. 3 how  data points are fitted by the 
Patterson et al.'s correlation (dashed line). 
The solid line  represents the evolutionary track of a 
1.0 ${M_\odot}$ secondary obtained using the two-dynamos evolutionary 
code developed by Zangrilli, Tout \& Bianchini (1997). 
The fit is particularly significant around and through the period gap, 
especially for filled circles. 
With the exception of V1309 Ori and AT Ara, the unstable 
systems candidates (labelled by crosses) that apparently possess 
unevolved secondaries do not follow any distinctive pattern and fall 
in the range $3<P_{orb}<6$ h, corresponding to masses of the 
secondaries in the range $0.3-0.7 M_\odot$. 
\smallskip

It is however possible that  mass transfer stability condition (2), 
also sketched in Fig. 1, is not  applicable to the secondaries 
of some CVs because their convective shells follow a mass-radius 
relation that implies larger $\xi_{ad}$ exponents.
Actually, the adopted stability criteria might be  uncertain, 
the masses of the two components could be  poorly determined, 
and  a significant fraction of short-period, angular momentum-driven 
CVs seem to have a somewhat evolved donor star (e.g. Kolb \& Baraffe 2000,
Kolb \& Williems 2005). 
\smallskip

In all cases, in the next paragraph we will focus our attention on 
the  role played by  the magnetic fields produced in the 
convective envelopes of the secondaries in changing 
the  $\xi_{ad}$ exponent and the critical $q_{dyn}$-profile.

\section{How  magnetic fields in the secondaries modify the 
critical q-profile of CVs}

It is widely recognized that  magnetic fields, increasing the 
critical Rayleigh number, inhibit convection (see, for example,
van den Borght 1969).
The possibility that convection inside stars might be suppressed
by  strong magnetic fields of simple geometry, namely, poloidal
and/or toroidal, was discussed by Gough \& Tayler (1966) and  
Moss \& Tayler (1969, 1970).
A detailed solar model containing a large-scale 
''magnetic perturbation''  to mimic the strong fields 
associated with an $\alpha-\Omega$-type solar dynamo  was constructed 
by  Lyndon \& Sofia (1995). They found that the changes in the 
temperatures and the adiabatic gradient produced by the magnetic 
perturbation lead to quite large decreases in the 
convective velocities above the perturbed region and to  considerable 
increases of  the convective turnover time.
\smallskip

Even neglecting  boundary-layer dynamos, which are 
responsible for the large scale structure of the stellar 
poloidal field,  chaotic envelope-dynamos 
driven by convective turbulence are sufficient to produce 
magnetic energy densities comparable to the kinetic energy, that is 
at a level close to global equipartition (Thelen \& Cattaneo 2000).
Liao \& Bi (2004) found that  turbulent
magnetic fields can also inhibit the generation of convection.
We might then think that if magnetic fields  can reduce the efficiency of 
convection,  they  should also reduce their adiabatic response 
to mass loss. 
However, since the prescription for dynamical stability is  
$\xi_{ad}~-~\xi_{RL2} > 0$, we should mainly investigate whether 
magnetic fields can modify the adiabatic mass-radius 
exponent $\xi_{ad}$.
\smallskip

For stars with extended convective layers, that means for masses
$\le0.5 M_\odot$, an approximated expression of the mass-radius 
adiabatic exponent given by Paczy\'nski (1965) is  
\begin{equation}
\xi_{ad}=\frac{\gamma-2}{3\gamma-4}
\end{equation}
where $\gamma = c_p/c_v$ is the polytropic exponent for an adiabatic
transformation and $c_p$ and $c_v$ are the specific heats.
The $\xi_{ad}$-vs-$\gamma$ diagram  is shown in Fig. 4.
For an ideal gas and  isotropic turbulence we have $\gamma = 5/3$ 
and $\xi_{ad}=-1/3$, which, in Fig. 1,  corresponds to the  
$q=0.63$ critital mass ratio  for the  dynamical stability 
of secondaries below  0.5$M_\odot$. 
From Fig. 4 we may see that, if we chose  $\gamma$ values
below 4/3 or above 5/3, we obtain  $\xi_{ad}>-1/3$ and thus  
steeper mass-radius relations that  lead to less restrictive 
critical $q-$profiles for low mass secondaries.  
\smallskip
 
The corrections to the  thermodynamical variables due to
the presence of a turbulent magnetic field derived by 
Liao \& Bi (2004) are
\begin{equation}
\Delta c_v= \beta_m \frac{k}{\mu m_u}
\end{equation}
\begin{equation}
\Delta c_p= 2\beta_m \frac{k}{\mu m_u}
\end{equation}
\begin{equation}
\Delta \gamma= \frac{1}{c_v}\Delta c_p - \frac{c_p}{c_v^2}\Delta c_v 
(2- \gamma) \frac{\Delta c_v}{c_v}
\end{equation}
where $\beta_m$ is the ratio of magnetic pressure to gas pressure,
while $k$, $\mu$ and $m_u$ have their usual meaning.
As we can see, a turbulent magnetic field tends to increase 
the specific heats and $\gamma$. Assuming an initial $\gamma=5/3$,
the increment of the polytropic exponent  can be written as
$\Delta \gamma\sim(0.33/c_v)\beta_m k/(\mu m_u)$. 
Fig. 5 plots the mass-radius exponent $\xi_{ad}$ derived
from eq. (4) as a function of $\beta_m$, for  complete 
and  partial hydrogen ionization, assuming unperturbed
specific heats $c_v/(\frac{3}{2}N_0k)=2$ 
and $c_v/(\frac{3}{2}N_0k)=35$, respectively (Clayton 1983,
table 2-4), $N_0$ being the Avogadro number. 
If we focus on  the full ionization line in Fig. 5 we find for 
$\beta_m=0.01$  an increase $\Delta \gamma/\gamma =1.6\times10^{-3}$, 
which is in good agreement with the numerical results obtained by 
Mollikutty, Das \& Tandon (1989) when we use the values of the 
specific heats listed in their Tables 1 and 2.
We conclude that chaotic dynamos can produce critical q-profiles 
above the minimum standard value $q_{dyn}=0.63$ for low mass secondaries 
(see Fig. 1), but probably not much above $q\sim0.7$ even for 
large $\beta_m$'s. This result shows that the 
modified critical q-profile  accounts  for the stability  
of the border line systems of Fig 1. 
\smallskip

If we consider large-scale, geometrically structured magnetic fields,
like those generated by $\alpha-\Omega$-type dynamos, then
the magnetic extra pressure  is no more a simple scalar and  
$\gamma$ will strongly depend on the field geometry.
Since the magnetic force component along the magnetic field is zero,
in  that direction a magnetically controlled gas will show $\gamma=1$, 
while, perpendicular to the field, it will be  
$\gamma=2$.   For fields tangled on distance scales of interest 
$\gamma=4/3$, as for a relativistic gas (Endal, Sofia \& Twigg 1985).
Thus, $\gamma$ represents the ratio of specific heats for the 
applied perturbation. 
In practice, the effective polytropic exponent of a convective envelope
should be calculated for any given field geometry, distribution and intensity
and averaged over the whole convective shell. 
In particular, as stellar structures mainly  depend  on the radial variations
of the total pressure, we will only consider  the radial dependence 
of the magnetic pressure and the polytropic exponent.
Lyndon \& Sofia (1995) studied the effects on the solar structure 
produced by localized large-scale magnetic fields with no radial component, 
i.e. a toroidal field with a radial pressure component,  
and assumed $\gamma=2$. 
Similarly, the  toroidal fields rising from the  overshoot regions
of secondaries with masses between 0.3 and 0.5 $M_\odot$ might also
produce strong localized magnetic perturbations with $5/3<\gamma<2$. 
Poloidal fields, instead, appear radial only at large 
latitudes and could, in principle, produce perturbations 
with  $1<\gamma<5/3$. We recall that radial fields should much more strongly 
inhibit convection (Moss \& Tayler 1969).
\smallskip

However, since  magnetic pressure is, on the average, a small 
fraction of gas pressure, the effective polytropic exponent of the 
convective shell should not greately differ from the  
unperturbed value  $\gamma=\frac{5}{3}$. 
Looking at Fig. 4, we see that, for  small changes of 
$\gamma$, the initial value of the adiabatic exponent  $\xi_{ad}=-\frac 13$ 
can   increase or decrease whether  the  geometry of the magnetic 
perturbation  corresponds to $\gamma=2$ or to $\gamma=1$, respectively. 
However,  the $\gamma=2$  geometry very likely   dominates
most of the stellar envelope  because large-scale poloidal fields are 
weaker than the toroidal ones and because they become radial only near 
the magnetic poles. 
Thus, we may roughly  describe  the effects of large scale 
fields using the same  approach as for turbulent fields,  
assuming $\beta_m$ as the ratio of the radial component of the combined
toroidal and poloidal magnetic pressure, $ B^2_{tot}/8\pi$, to gas pressure.
According to Applegate (1992), the  magnetic torques which are active in 
the outer convective regions of the secondaries should be produced by
fields of several kilogauss. Toroidal  fields of some $10^4$ G
produced by $\alpha-\omega$ dynamo models were suggested by Zangrilli, 
Tout \& Bianchini (1997) in CVs just above the period gap.  
Since  equipartition, i.e. $\beta_m=1$, in the envelopes of low mass 
secondaries typically requires fields of $\sim10^8$ G, the ratio 
of magnetic to gas pressure is  $\beta_m\sim 10^{-4}$.
This small value  however becomes  $\sim 100$ times larger 
at the  photosphere (Applegate \& Patterson 1987).
Assuming   $\beta_m\sim 10^{-4}$,  equations 5-7 yield  
$\Delta\gamma\sim 3\times10^{-5}$, while from eq. 4 we obtain
$\Delta\xi_{ad}\sim5\times10^{-5}$.
We  conclude that  large-scale toroidal and poloidal  
fields should not significantly contribute to increase the stability of
fully, or almost fully convective secondaries, their effects being  smaller
than those produced by chaotic fields. 
\smallskip

Possible effects due to the  presence of magnetic torques in 
the outer convective regions should also be investigated.
 Applegate (1992) explained that the field lines rising from 
the magnetic dynamo  in the overshoot region produce a negative 
torque in the outer envelope trying 
to spin down the tidally locked outer layers of the secondary. 
In this case, an observer corotating with the binary would see 
the surface of the star rotate in a retrograde sense. 
The consequent change in the quadrupole moment of the star 
would then produce the orbital period modulations observed 
in some CVs during magnetic cycles. 
Richman, Applegate \& Patterson (1994)  have 
shown that a consequence of the  decreased rotation of the outer 
envelope is the appearence of an inward directed Coriolis acceleration. 
Following these authors, we find that for orbital periods  
around $4$ h and  $\sim 0.45$ $M\odot$ secondaries, the 
mean  extra-gravity due to Coriolis forces is 
$\sim 10^2~~ cm~ s^{-2}$. 
Since this is  only  $10^{-3}$ times the  surface  gravity,
some effects could be seen only in the photosphere and the 
thin radiative layer above.
Another consequence is that the energy stored and dissipated 
in the  convective zone, where a peak of differential rotation 
is cyclically set up by  magnetic cycles, will produce a modulation
of the stellar luminosity  of the order of  $\Delta L/L\sim 0.1$,
corresponding to   changes in the effective temperature  
$\Delta T_{eff}/T_{eff}\sim 0.025$.
In Applegate's model, the release of the stored torque 
energy should heat not only the stellar surface, but most of 
the convective shell as well.
Since  in most of the convective region the temperatures are above 
the HeII ionization limit, the Rosseland mean is dominated 
by Kramers' law. Thus, inside the convective shell, a $3\%$ 
increment of the temperature should  produce a $\sim 10\%$ 
decrease in the opacity coefficient. The result should be a less 
developed convection zone and a hotter photosphere, the effect being 
more pronounced as the mass of the secondary increases 
(Pizzolato et al. 2001).
Instead, in the cool atmospheres, small temperature increments 
should determine substantial increases of the mean opacity 
and the atmospheric radiative gradient.
As a result, the photospheric radius should  increase.
However, whether the combined effects of  all these mechanisms will 
result in a steeper mass-radius relation with a 
larger $\xi_{ad}$ could be understood only by performing 
numerical simulations with full stellar models.

\section{The effects of magnetic cycles on the mass transfer rate}

Since  $\alpha-\Omega$-type dynamos are 
characterized by periodic solutions (magnetic cycles), we wonder 
whether the presence of  weakly variable $\gamma$ and $\xi_{ad}$ might  
represent an additional/alternative mechanism  
modulating the mass transfer rate within the binary system.
Following Osaki (1985), Warner (1988) and Richman, Applegate \& Patterson
(1994), we may express the fractional change in the accretion rate as 
\begin{equation}
\Delta\dot M_2/\dot M_2 = \Delta R/H 
\end{equation}
where $H$ is the  photospheric pressure scale height that, for a lower 
main sequence star, is $\sim3200$ times smaller than the radius $R$, 
and $\Delta R$ is here assumed as the variation of the
stellar radius produced by  a change of $\xi_{ad}$.
We may then rewrite  expression  8  as 
\begin{equation}
\Delta\dot M_2/\dot M_2 \sim 3200 \times \Delta R/R 
\end{equation}
and, since 
\begin{equation}
\Delta R/R \sim lnM\Delta\xi_{ad}
\end{equation}
we obtain, for a 0.5 $M_\odot$ secondary,  $\Delta\dot M_2/\dot M_2 \sim 0.1$.
Actually, this is  the order of magnitude of the observed long-term luminosity 
variations of CVs which are usually ascribed to the presence in the 
secondaries of solar-type cycles (Bianchini 1987, 1990; Warner 1988;
Richman, Applegate \& Patterson 1994).

\section{Conclusions }
A few CVs  with  orbital periods in the range 3.5-7 h have
mass ratios that would define them as thermally stable 
but dynamically unstable (Fig. 1). Assuming that the masses of the 
two companions are fairly well determined and the secondaries are MS stars, 
the commonly adopted criteria do not explain the stability of these 
systems against mass tansfer.
However, we found another effect since  cahotic and large-scale   
magnetic fields  produced in the convective shells of the secondaries  
modify the polytropic $\gamma$ exponent,
the  mass-radius adiabatic exponent $\xi_{ad}$
and, consequently,  the critical $q$-profile for stable 
mass transfer.
We demonstrated that  large $\beta_m$'s turbulent magnetic 
fields might produce critical q-profiles  10\%  higher than the 
usually adopted minimum critical value $q_{dyn}=0.63$ for 
low mass secondaries. 
We found that the contribution by large-scale  toroidal and poloidal 
fields could be smaller.  Thus, magnetic fields alone
cannot  account for the stability of all the anomalous CVs 
identified in Fig. 1 and other explanations must be investigated.

During this study, we found another important effect due to the magnetic 
fields of the secondaries. In fact, since toroidal and poloidal fields are 
produced by  $\alpha-\Omega$-type dynamos, their intensities 
should vary throughout magnetic cycles, inducing periodic changes in 
the mass-radius adiabatic exponent $\xi_{ad}$. 
We have demonstrated that these variations, though  small,
are sufficient to explain the cyclic mass transfer rate  
variations observed in a number of CVs.

\section*{ACKNOWLEDGEMENTS}
A.B. wishes to thank the people of the Department of Physics and  
Astronomy of the University of Wyoming for their friendship and support.
We acknowledge Cesare Chiosi, Franca D'Antona, Alvio Renzini and 
Marina Orio for helpful discussions. 
This research was supported by NASA Cooperative Agreement NCC 5-578 to
the Wyoming EPSCoR Program  and by the Italian MURST.

\clearpage

\begin{figure}
\includegraphics[clip=,width=0.8\textwidth]{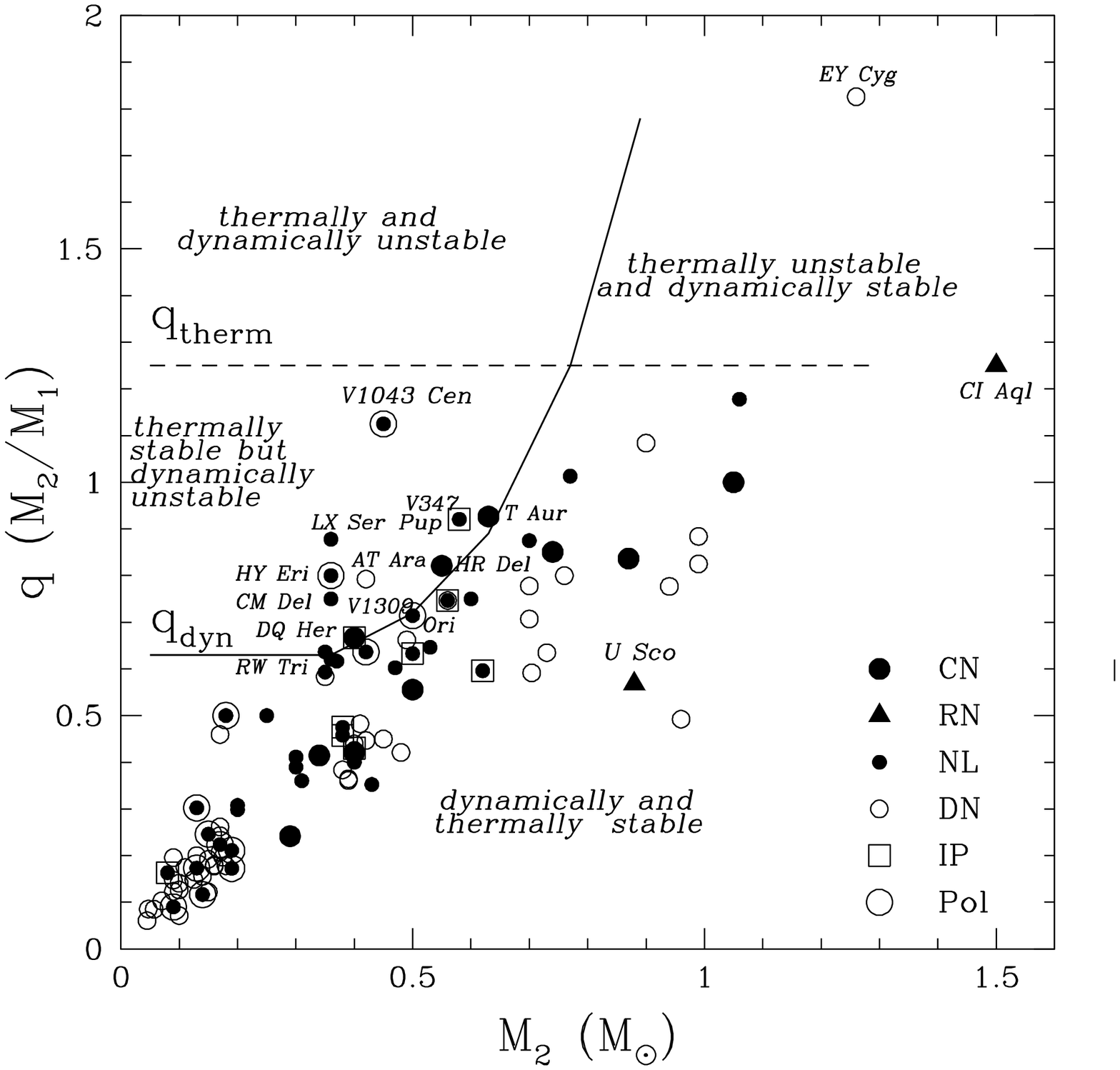}
\caption{Plot of  mass ratios versus the masses of the secondaries
(from RK2005). {\it Open circles} represent dwarf novae;
{\it small solid circles} represent nova-like systems;
{\it large solid circles} are classical novae;
the {\it solid triangles} are
recurrent novae. Symbols surrounded by {\it open squares}
are intermediate polars while those surrounded by {\it open circles}
are polars. The critical line for dynamical stability $q_{dyn}$ (solid line),
and that for  thermal stability  $q_{therm}$ (dashed line), are shown.
Six objects fall inside the thermally stable but dynamically unstable
region (see text). Their error bars are shown in Fig. 2.
}
\label{stab}
\end{figure}

\clearpage

\begin{figure}
\includegraphics[clip=,width=0.8\textwidth]{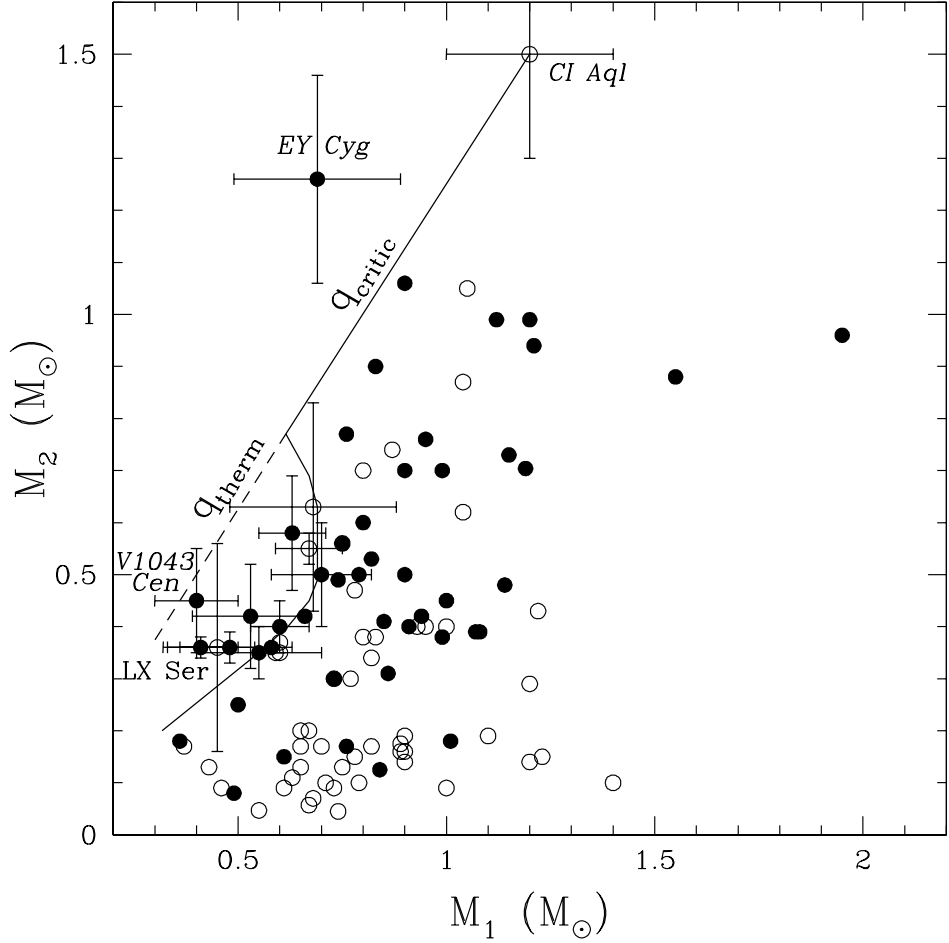}
\caption{Plot of the masses of the components.
Here, {\it filled circles} refer to  spectroscopic determinations, 
while {\it open circles} mean that the mass of the secondary 
was derived from empirical correlations (namely, the orbital period). 
The $q_{critic}$ line is shown by a solid line; the dashed line 
shows  the upper  limit for  thermal stability.    
The unstable systems are plotted with their error bars.  
Most of them tend to  cluster  around  $M_1\sim 0.55 M_\odot$ and 
$M_2\sim 0.38 M_\odot$ (see text). V1043 Cen and LX Ser are 
unstable at a 95$\%$  confidence level.
 }
\label{stab}
\end{figure} 

\clearpage

\begin{figure} 
\includegraphics[clip=,width=0.8\textwidth]{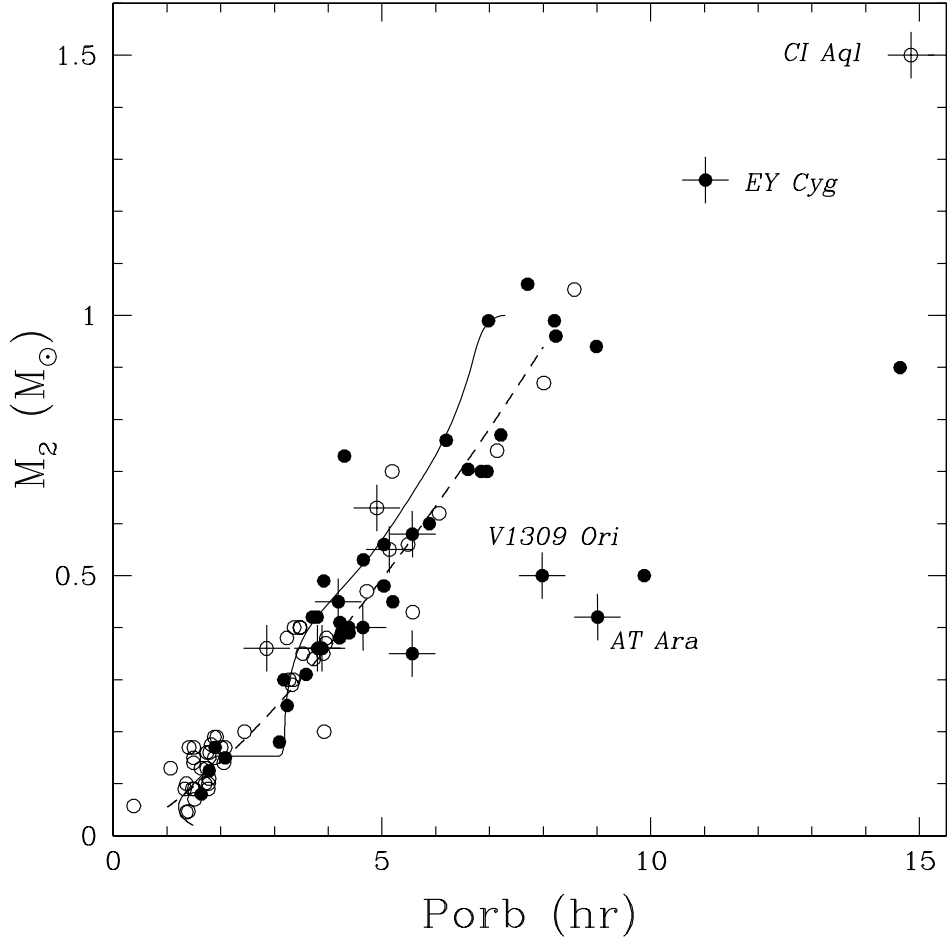}  
\caption{The $M_2-P_{orb}$ correlation. The dashed line represents
the empirical law by Patterson et al. (2003). The solid line is 
the evolutionary track of a 1.0 ${M_\odot}$ secondary. 
Symbols are like in Fig. 3.
In this diagram, with the exception of  V1309 Ori, AT Ara and
the long-period recurrent nova CI Aql, unstable objects, 
marked by crosses, do not show any distinctive pattern.
   } 
\label{stab} 
\end{figure} 

\clearpage

\begin{figure} 
\includegraphics[clip=,width=0.6\textwidth]{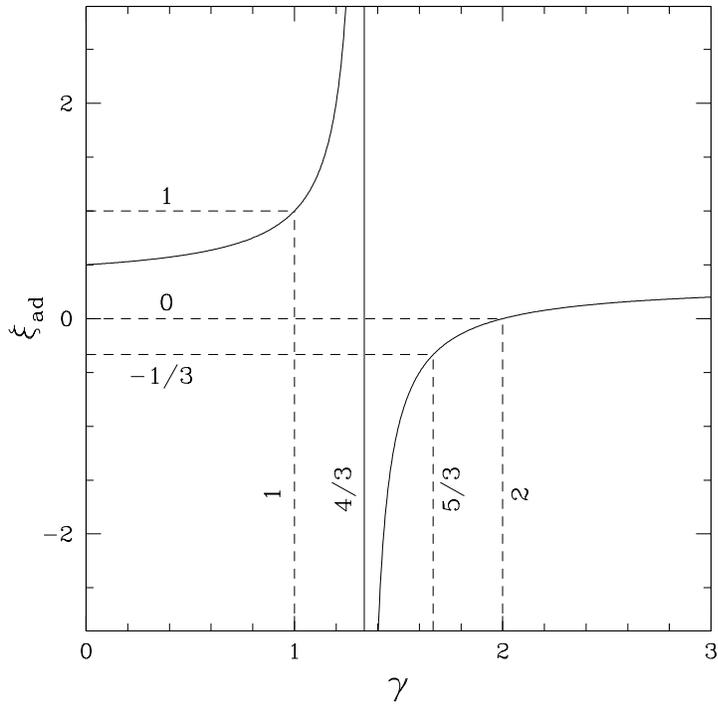}  
\caption{ 
The $\xi_{ad}-\gamma$ diagram for secondaries with extended
convective envelopes, namely, with masses $\le0.5M_\odot$.
  } 
\label{xigam} 
\end{figure} 

\clearpage

\begin{figure} 
\includegraphics[clip=,width=0.6\textwidth]{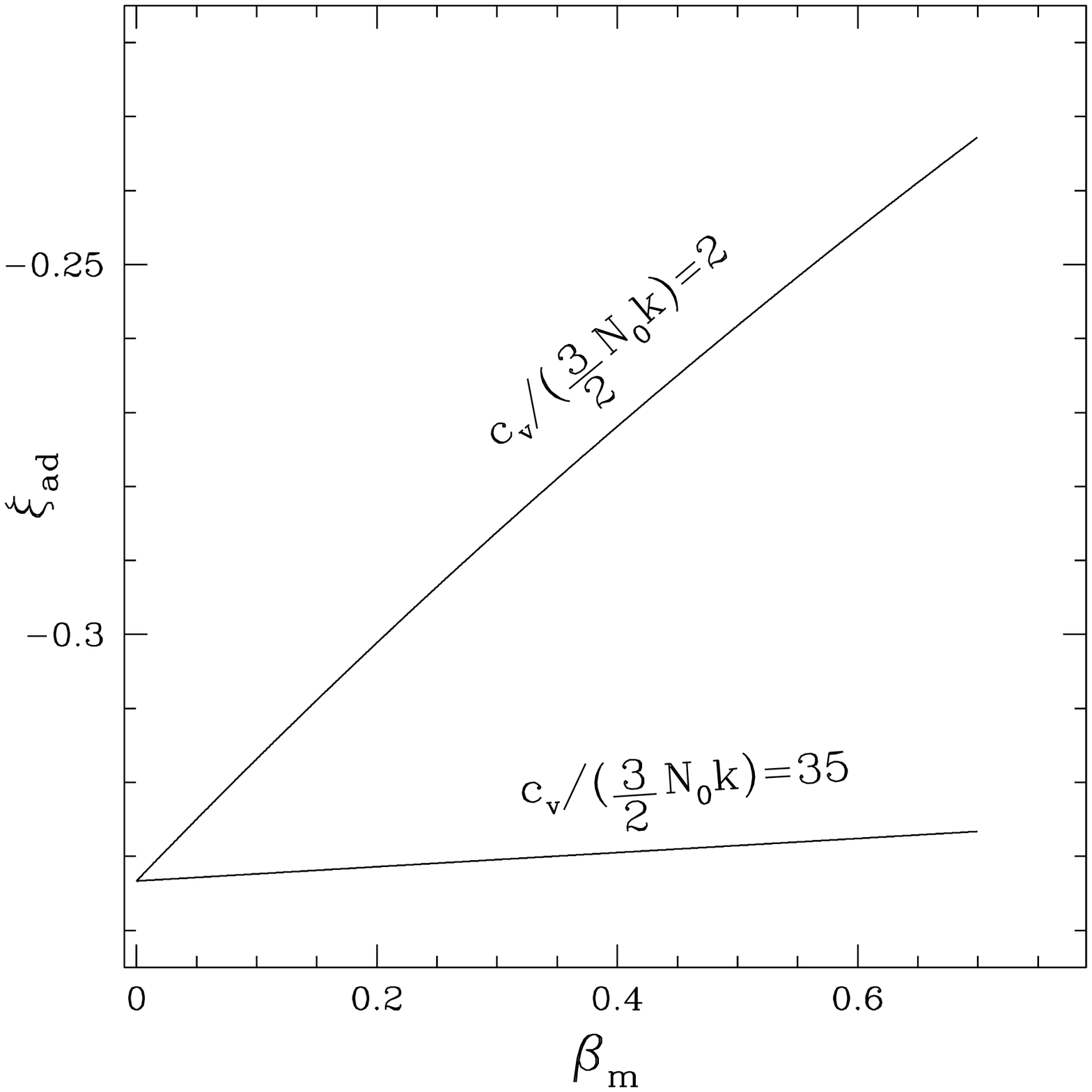}  
\caption{The mass-radius adiabatic exponent $\xi_{ad}$  
as a function of $\beta_m$ for turbulent magnetic fields. 
The cases of complete and  
partial (50\%) hydrogen ionization are labelled by the 
values $c_v/(\frac{3}{2}N_0k)=2$ 
and $c_v/(\frac{3}{2}N_0k)=35$, respectively.
   } 
\label{stab} 
\end{figure}


\begin{thebibliography}{}

\bibitem{applegate} Applegate, J. H., 1992, ApJ, 385, 621.

\bibitem{applegate87}Applegate, J. H., Patterson, J., 1987, ApJ, 322, L99.
  
\bibitem{baptista} Baptista, R., Horne, K., Hilditch, R.W., 
Mason, K.O., Drew, J.H., 1995, ApJ, 448, 395.

\bibitem{baraffe} Baraffe, I., Kolb, U., 2000, MNRAS, 318, 354.

\bibitem{bianc87}Bianchini, A., 1987, Mem. SAI, 58, 245. 
 
\bibitem{bianc90}Bianchini, A., 1990, AJ, 99, 1941.

\bibitem{clayton} Clayton, D. D., 1983, Principles of Stellar
Evolution and Nucleosynthesis, (Chicago: Univ. Chicago Press).

\bibitem{endal} Endal, A. S., Sabatino, Sofia, S., \& Twigg, L. W.,
1985, ApJ, 290, 748. 

\bibitem{dekool92} de Kool, M., 1992, A\&A, 261, 188.

\bibitem{gansike2003} G\"ansike, B.T., Szkody, P., De Martino, D., 
Beuermann, K., Long, K.S., Sion, E.M., Knigge, C., Marsh, T., 
Hubeny, I., 2003, ApJ, 594, 443.

\bibitem{gough} Gough, D. O., \& Tayler, R. J., 1966, MNRAS, 133,85. 

\bibitem{hall1991} Hall, D.S., 1991, ApJ, 380, L85.

\bibitem{howell}Howell, S.H. \& Skidmore,W.,2000, New A. Rev.,44,33.

\bibitem{king94} King, A.R., Kolb, U., de Kool, M., \& Ritter, H., 
1994, MNRAS, 269, 907.

\bibitem{king96} King, A. R., Frank, J., Kolb, U., Ritter, H., 1996,
ApJ, 467, 761.
 
\bibitem{kolb} Kolb, U., Baraffe, 2000, New Astronomy Rev., 44, 99.

\bibitem{kolbwilliems} Kolb, U., Williems, B., 2005, ``The Astrophysics 
of Cataclysmic Variables and Related Objects'', Proceedings of ASP 
Conference Vol. 330. Edited by J.-M. Hameury and J.-P. Lasota. 
San Francisco: Astronomical Society of the Pacific, 2005., p.17


\bibitem{lederle} Lederle, C.,\& Kimeswenger, S., 2003, A\&A, 397, 951L.

\bibitem{liao2004} Liao, Y. \& Bi, S. L., B., 2004, Ch. Astron. Astrophys.,
4, 490.

\bibitem{livio92}Livio, M., 1992, ApJ, 393, 516.

\bibitem{lyndon} Lyndon, T. J., \& Sofia, S., 1995, 
ApJ {\it Suppl. Series}, 101, 357.

\bibitem{marsh} Marsh, T. R., \& Pringle, J. E., 1990, ApJ, 365, 677.

\bibitem{mollikutty} Mollikutty, O.J., Das, M.K., \& Tandon, J.N.,
1989, Astrophys. Space Sci., 155, 249. 

\bibitem{moss69} Moss, D. L., \& Tayler, R. J., 1969, MNRAS, 145, 217.

\bibitem{moss70} Moss, D. L., \& Tayler, R. J., 1970, MNRAS, 147, 133.

\bibitem{osaki} Osaki, Y., 1985, A\&A, 144, 369.

\bibitem{paczynski65} Paczy\'nski, B., 1965, Acta Astron., 15, 89.

\bibitem{paczynski76} Paczy\'nski, B., 1976, in The Structure and Evolution 
of Close Binary Systems, eds. P. Eggleton, S. Mitton and J. Whelan,
Dordrecht:Reidel, 75. 

\bibitem{patterson2003} Patterson, J., Thorstensen, J. R., Kemp, J.,
Skillman, D. R., Vanmunster, T., Harvey, D. A.,  and 20 more authors,  
2003, PASP 115, 1308.

\bibitem{pizzolato} Pizzolato, N., Ventura, P., D'Antona, F., Maggio, A., 
Micela, G., Sciortino, S., 2001, A\&A, 373, 597.

\bibitem{Putte2003} Putte, D. V., Smith, R. C., Hawkins, N. A. \& 
Martin, J. S., 2003, MNRAS, 342, 151.

\bibitem{rappaport} Rappaport, S., Verbunt, F., \& Joss, P. C., 1983,
ApJ, 275, 713.

\bibitem{richman} Richman, H. R., Applegate, J. H., \& Patterson, J.,
1994, PASP, 106, 1075.

\bibitem{RK} Ritter, H. \& Kolb, U., 2005,
http://vizier.cfa.harvard.edu/viz-bin/VizieR.


\bibitem{schenker}Schenker, K., Kolb, U., \& Ritter, H., 1998, MNRAS,297,633.

\bibitem{shara} Shara, M. M., Livio, M., Moffat, A. F. J., Orio, M., 1986,
ApJ, 311, 163.

\bibitem{sion} Sion, E.M., Winter, L., Urban, J.A., Tovmassian, G.H., 
Zarikov, S., G\"aniscke, B.T., Orio, M., 2004, AJ, 128, 1795.

\bibitem{smith} Smith, D. A. \& Dhillon, V. S., 1998, MNRAS, 301, 767. 

\bibitem{spruit} Spruit, H. C., \& Ritter, H., 1983, A\&A, 124, 267.

\bibitem{stehle} Stehle, R., Ritter, H., Kolb, U., 1996, MNRAS, 279, 581. 

\bibitem{thelen} Thelen, J.-C, \& Cattaneo, F., 2000, MNRAS, 315, L13.

\bibitem{tovmassian} Tovmassian, G., Orio, M., Zharikov, S., Echevarria,
J., Costero, R., Michel, R., 2002, in AIP Conf. Proc. 637, 
{\it Classical Nova Explosion}, ed. M. Hernanz \& J. Jos\`e 
(Melville:AIP),72.

\bibitem{vandenborght} van den Borght, R., 1969, A\&A, 2, 96.

\bibitem{warner88} Warner, B., 1988, Nature, 336,129.   
     
\bibitem{warner95a} Warner, B.,1995a, Cataclysmic Variable Stars, 
Cambridge Univ. Press, Cambridge.
 
\bibitem{warner95b} Warner, B.,1995b, Ap\&SS, 232, 89.

\bibitem{whyte} Whyte, C., Eggleton, P.P., 1980, MNRAS, 190, 801.

\bibitem{zangrilli} Zangrilli, L., Tout, C. A., \& Bianchini, A., 1997
MNRAS, 289, 59.

\end{thebibliography}
\end{document}